\documentclass[lettersize,journal]{IEEEtran}
\usepackage{amsmath,amsfonts}
\usepackage{algorithmic}
\usepackage{algorithm}
\usepackage{array}
\usepackage[caption=false,font=normalsize,labelfont=sf,textfont=sf]{subfig}
\usepackage{textcomp}
\usepackage{stfloats}
\usepackage{url}
\usepackage{verbatim}
\usepackage{graphicx}
\usepackage{cite}
\usepackage[colorlinks, citecolor=blue]{hyperref}
\usepackage [english]{babel}
\usepackage [autostyle, english = american]{csquotes}
\MakeOuterQuote{"}
\begin{document}

\title{Synergistic Interplay of Large Language Model and Digital Twin for Autonomous Optical Networks: Field Demonstrations}

\author{Yuchen Song, Yao Zhang, Anni Zhou, Yan Shi, Shikui Shen, Xiongyan Tang, Jin Li, Min Zhang, Danshi Wang,~\IEEEmembership{Senior Member,~IEEE,}

\thanks{This work was supported by National Natural Science Foundation of China No. 62171053, Beijing Nova Program No. 20230484331 (Corresponding author: Danshi Wang).}

}

\markboth{Accepted by IEEE COmmunications Magazine, 2024}%
{Shell \MakeLowercase{\textit{et al.}}: A Sample Article Using IEEEtran.cls for IEEE Journals}


\maketitle

\begin{abstract}
The development of large language models (LLM) has revolutionized various fields and is anticipated to drive the advancement of autonomous systems. In the context of autonomous optical networks, creating a high-level cognitive agent in the control layer remains a challenge. However, LLM is primarily developed for natural language processing tasks, rendering them less effective in predicting the physical dynamics of optical communications. Moreover, optical networks demand rigorous stability, where direct deployment of strategies generated from LLM poses safety concerns. In this paper, a digital twin (DT)-enhanced LLM scheme is proposed to facilitate autonomous optical networks. By leveraging monitoring data and advanced models, the DT of optical networks can accurately characterize their physical dynamics, furnishing LLMs with dynamic-updated information for reliable decision-making. Prior to deployment, the generated strategies from LLM can be pre-verified in the DT platform, which also provides feedback to the LLM for further refinement of strategies. The synergistic interplay between DT and LLM for autonomous optical networks is demonstrated through three scenarios: performance optimization under dynamic loadings in an experimental C+L-band long-haul transmission link, protection switching for device upgrading in a field-deployed six-node mesh network, and performance recovery after fiber cuts in a field-deployed C+L-band transmission link.
\end{abstract}


\section{Introduction}
The rapid development and extensive cross-applications of artificial intelligence (AI) have propelled the evolution of optical networks from a static and manual mode to an autonomous one \cite{musumeci2018overview}. However, previous studies have mainly relied on conventional machine learning or deep learning models with relatively small sizes, which exhibited limited intelligence and could only fulfill one or two specific tasks, but always falling short of multi-task implementations and far from full automation capabilities \cite{zheng2023automation}. Moreover, human involvement is still required, especially in invoking these techniques and conducting data analysis to proceed to the next step in the workflow. Central to the realization of autonomous optical networks is to create a versatile "AI Agent" in the control plane, which can perform comprehensive management systematically and execute various tasks methodically for autonomous operation. Currently, large language models (LLMs) as generative AI techniques have sparked a revolution in all walks of life \cite{min2023recent}. Typically, LLMs are generalist models that can effectively solve general-purpose natural language processing (NLP) tasks on a broad scale, but it still faces challenges when tackling specific or complex tasks within specific professional domains \cite{lee2023benefits}. 

The remarkable advancement and powerful capabilities of LLMs prompt the community to expect the prospects of autonomous networks \cite{zhang2024gpt}. Nevertheless, integrating the LLM with optical networks presents several challenges. First, the behaviors of optical networks involve various nonlinear dynamics, which are characterized by mathematical and physical laws (e.g., nonlinear fiber optics). However, general LLMs, originally designed for NLP tasks, are incapable in deeply understanding and accurately describing the intricate behaviors of optical transmission \cite{jiang2024opticomm}. For instance, LLMs cannot efficiently solve the nonlinear Schrödinger equation (NLSE), which governs fiber channel behavior. This mathematical limitation significantly hinders the application of LLMs in optical networks, as they cannot fully comprehend or predict network dynamics. Secondly, LLMs is expected to access streaming data from optical networks, including device configurations, node powers, signal routes, and transmission performance metrics. This data is crucial during the early stages of prompt engineering or fine-tuning for LLMs and is essential for deriving timely and accurate strategies. How to establish such efficient data transmission under standardized protocols remains a challenge. Third, considering the pivotal role of optical networks in supporting global internet and big data transmission, it is paramount to ensure rigorous stability and mitigate any risks of incorrect operation. Therefore, the derived management strategies of LLMs necessitate thorough preview for performance assessment before they are deployed on field networks. In order to elicit LLM’s proficiency for optical networks, it is imperative to formulate effective prompt strategies, establish specialized knowledge bases, and develop appropriate augmented tools.

Recently, digital twin (DT) have been widely studied for optical networks to serve multiple purpose including monitoring current operational states, predicting future behavioral patterns, and estimating quality of transmission (QoT) \cite{zhuge2023building}. The DT of optical network is expected to respond to "what-if" scenarios involving planning, troubleshooting, upgrading, and other proactive analytics \cite{wang2024digital}, which can be used for strategy verification. Benefiting from these advantages, DTs are poised to enhance the capabilities of LLMs for autonomous optical networks in aiding decision-making processes and verifying generated strategies.

In this paper, we first implement accurate and dynamic-updating DT in field-trial optical networks with hybrid data-driven and physics-informed approach. The accurately updated DT can provide essential information for decision-making processes within LLMs and serve as a platform for previewing strategies generated by LLMs. In this case study, the advanced Generative Pre-trained Transformer-4 (GPT-4) is selected as the engine of AI Agent in controller of optical networks. We integrate domain knowledge into the LLM through prompt engineering and leverage external plugins and tools to facilitate management of optical networks. The synergistic interplay between LLM and DT is demonstrated through three field-trial optical transmission systems ranging from experimental C+L-band long-haul transmission link to field-deployed mesh networks, including scenarios of dynamic loadings, protection switching, and fiber cuts. 

The rest of the paper is organized as follows. In Section II, we introduce the framework for the interplay between LLM and DT, where DT collects data from optical networks and provide it to the LLM for further processing. The strategies generated by the LLM are then pushed to the DT for verification and, if deemed effective and safe, can be further implemented in the optical networks. Section III discusses the establishment of DTs using a hybrid data-driven and physics-informed deep learning approach, along with results on parameter refinement from three deployed optical transmission systems. In Section IV, we delve into the LLM-empowered AI agent, detailing its domain knowledge through prompt engineering and its use of external tools via API integration. Section V demonstrates the interaction between the LLM-empowered AI agent and DTs on the three deployed systems. Finally, conclusions are drawn in Section VI.



\section{Framework: Interplay between LLM and DT for autonomous optical networks}
\begin{figure}[t]
\centering\includegraphics[width=0.48\textwidth]{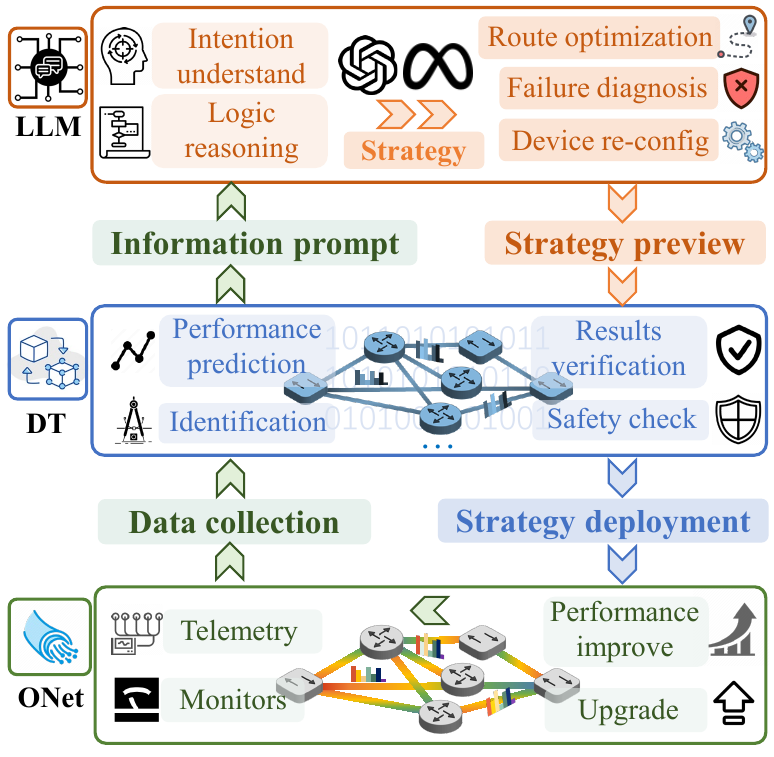}
\caption{Schematic of the synergistic interplay between DT of optical networks (ONet) and LLMs.}
\label{fig1}
\end{figure}

The framework of interplay between DT and LLMs for autonomous optical networks is illustrated in Fig. \ref{fig1}. DT is expected to use mirroring models, monitoring information, and data transfer mechanisms to depict and predict the activities of optical networks over their lifetime. To establish accurate DT of optical networks, the first step is to obtain precise values of their physical parameters. However, for field-deployed optical networks in practical environments, most parameters deviate from the nominal values provided by point-of-manufacture handbooks, and unexpected human activities and device ageing incur problems of parameter shifting \cite{song2023implementing}. To enable accurate establishment and dynamic updating of DT with collected monitoring data, we propose a hybrid data-driven and physics-informed approach for modeling the fiber channel within the DT. This technique aims to match the collected input and output channel power profiles from the optical channel monitor (OCM) by adjusting parameters within physical laws. This approach allows for the identification of key parameters along the transmission link, such as connector loss, fiber Raman gain strength, and amplifier gain spectrum. This process enables the establishment of an accurately updated DT, which serves as the foundation for subsequent management via LLM.

The DT possesses the capability to simulate and predict network behaviors, complementing the role of NLP-based LLMs, which excel primarily in intention analysis, semantic analysis, and logic reasoning. In autonomous optical networks, the DT continuously reports information to the LLMs, which in turn offers corresponding suggestions and strategies. LLMs based on transformer architecture and multi-head attention mechanisms can be prompted or fine-tuned to generate effective strategies for optical networks. Then the generated strategies undergo thorough previewing and verification in the DT platform, ensuring that only safe and certified strategies are deployed to physical network, as shown in Fig. \ref{fig1}. Additionally, verification conducted by the DT can provide feedback to the LLM for further refinement of strategies. This iterative process fosters a collaborative framework for autonomous management, leveraging the strengths of both DT and LLM while upholding high efficiency and safety standards.

\begin{figure*}[t]
\centering\includegraphics[width=1\textwidth]{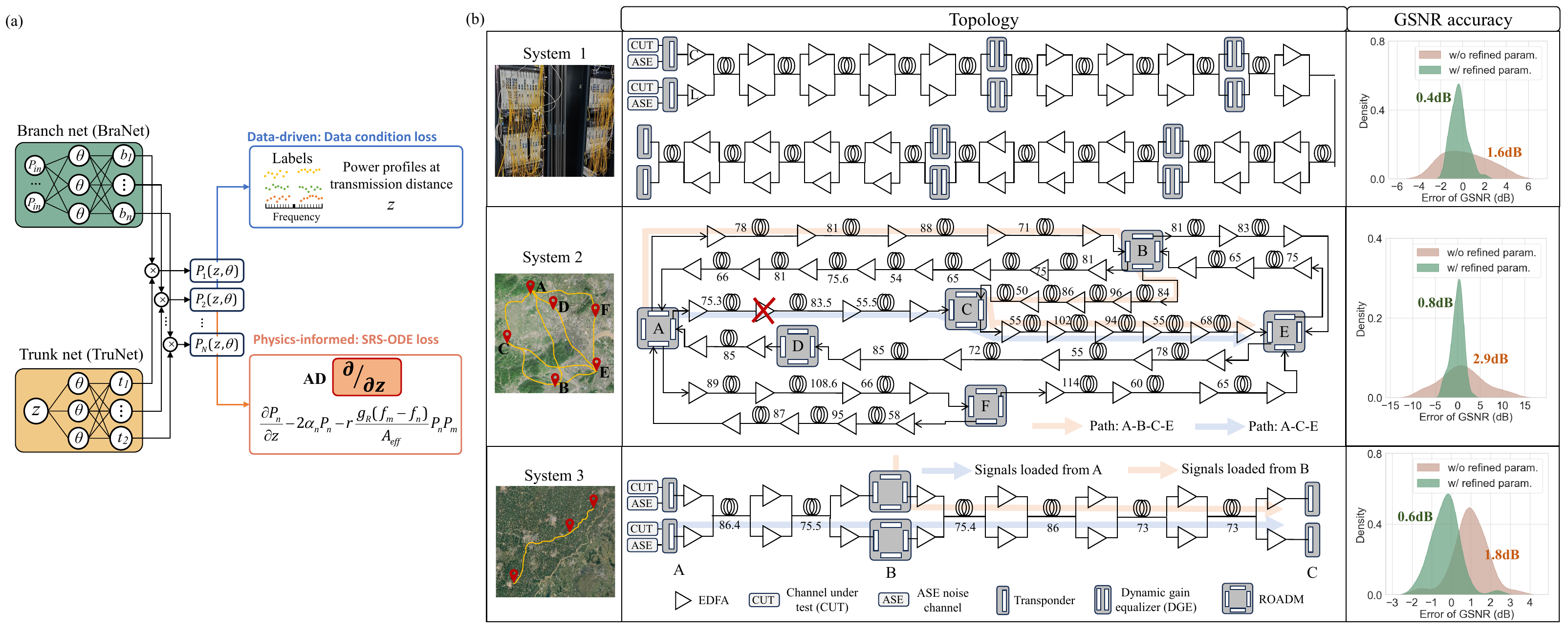}
\caption{Schematic of (a) hybrid data-driven and physics-informed DeepONet for fiber multi-channel power evolution modeling in DT. The topology of a long-haul C+L-band experimental link, a field-deployed six-node mesh network, and a field-deployed C+L-band link is illustrated in (b) with established DT GSNR accuracy calculated by MSE with positive for aggressive and negative for conservative. Mean absolute error is marked for GSNR accuracy w/ and w/o DT refined physical parameters.}
\label{fig2}
\end{figure*}

\section{Dynamic-updating DT of optical networks}
\subsection{Principles of hybrid data-driven and physics-informed deep operator network for fiber channel modeling in DT}
In the digital twin (DT) of optical networks, the most critical component is the modeling of the fiber channel. Based on our previous work, the fiber channel is modeled using deep operator networks (DeepONet) \cite{lu2021learning} in a hybrid data-driven and physics-informed approach. The structure of DeepONet comprises the branch net (BraNet) and the trunk net (TruNet), as shown in Fig. \ref{fig2}(a). The TruNet samples the transmission distance $z$ as inputs while the BraNet takes initial channel power profiles as input. The DeepONet outputs power profiles $P_n(z, \theta)$ at $z$. The training of DeepONet can be trained by the regularization of stimulated Raman scattering (SRS)-ordinary differential equations (ODEs) as well as collected labels at corresponding transmission distances. Further details on training DeepONet for fiber channel modeling can be found in \cite{song2023physics}.

\begin{figure*}[t]
\centering\includegraphics[width=0.95\textwidth]{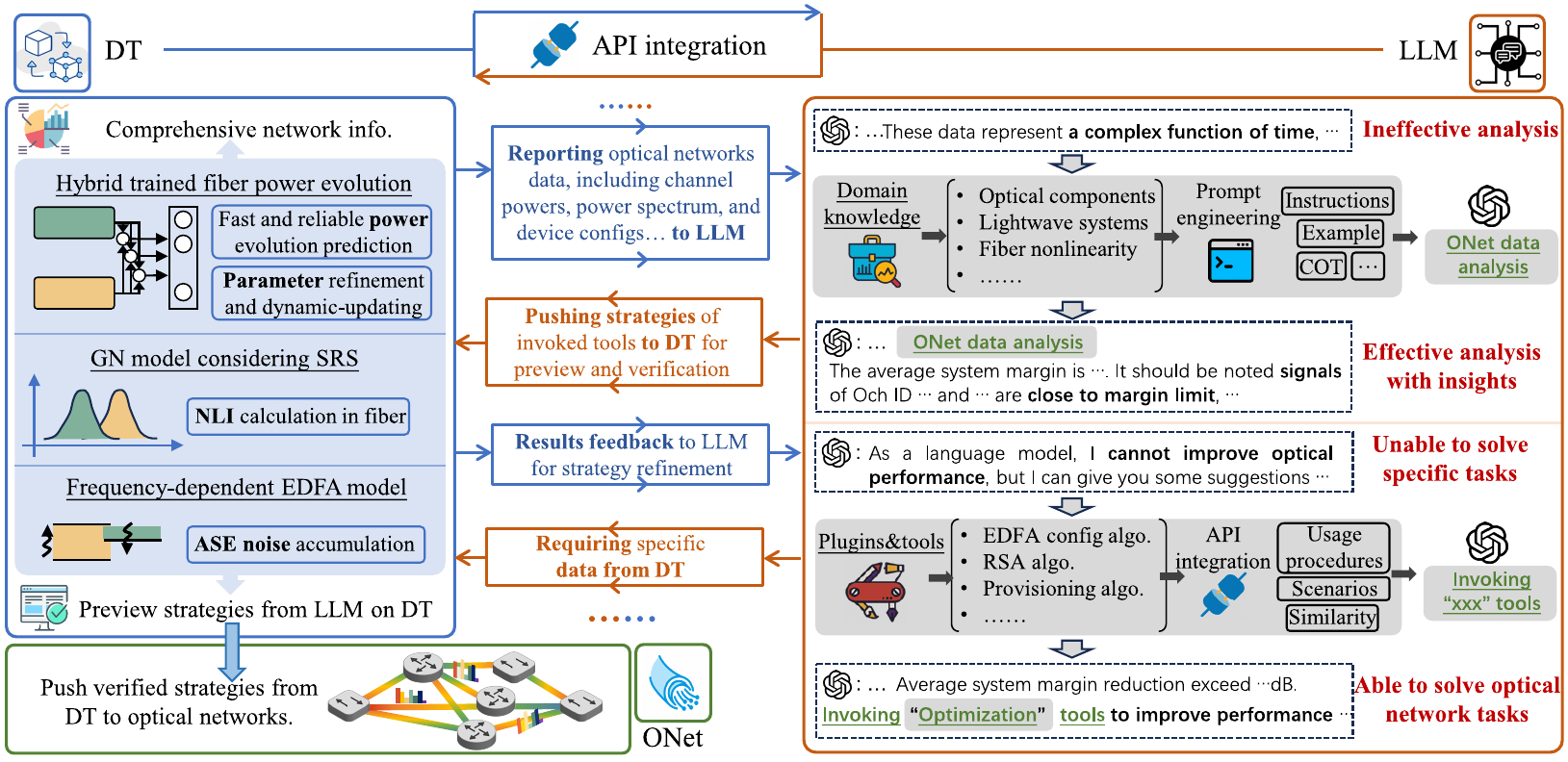}
\caption{Schematic of the interplay between DT, including functions of power evolution prediction, parameter refinement, NLI calculation and ASE noise accumulation, and LLM, equipped with domain knowledge and plugins and tools. Verified strategies are pushed from the DT to ONet.}
\label{fig3}
\end{figure*}

For forward performance estimation, the trained fiber channel model can be used to predict transmitted power profiles with strong generalization ability. The generalized signal-to-noise ratio (GSNR), which considers both the amplified spontaneous emission (ASE) noise from amplifiers and nonlinear interference (NLI) from fibers, is used as QoT metric. In this case, the NLI can be derived from Gaussian noise (GN) model considering SRS \cite{semrau2019closed}, and ASE noise can be calculated by frequency-dependent erbium-doped optical fiber amplifier (EDFA) model, as shown in Fig. \ref{fig3}. Moreover, to establish accurate and dynamic-updating DT of field-deployed networks, the critical physical parameters can be refined by re-training the trained DeepONet with collected power profiles by OCM at $z=0$ and $z_{max}$ of a span \cite{song2024srs}.

\subsection{Implementing DT on field-trial systems}
To demonstrate the feasibility of proposed DT technique, three practical systems are set up for test as illustrated in Fig. \ref{fig2}(d). In System 1, the laboratory C32+L32 WDM long-haul transmission link consists of 22 spans with 100km G.654 SMF in each one. The transmission bandwidth occupies the L-band, from 186.15 THz to 190.8 THz, and the C-band, from 191.35 THz to 196 THz with a total of 64 channels and 150 GHz channel bandwidth. Six commercial 400Gb/s transponders are configured for channels under test (CUT), and the rest of bandwidth can be filled with ASE noise channel. 

In System 2, the field-deployed mesh optical network consists of 6 ROADM sites. Each span ranges from 47.0 km to 114.0 km using G.652 SMF. Each transmitter site is equipped with the same commercial 400Gb/s transponders (up to 12 CUT) for a total of 60 channels with 100 GHz spacing cover the whole C-band. No ASE channel is used in this system.

In System 3, the field-trial C48+L48 WDM transmission link consists of six amplified spans with a maximum length of 86.4km (totaling 469.3km of G.652 SMF). The transmission bandwidth occupies the L-band, from 186.1 THz to 190.8 THz, and the C-band, from 191.4 THz to 196.1 THz with a total of 96 channels. Five commercial transponders are configured for five CUT, and PCS-16QAM with 91.6 baud rate is modulated with 100GHz channel spacing. Other channels can be filled with filtered ASE noise channel.

With collected channel powers by OCM, the DT is calibrated with actual optical networks and been verified under different loadings. The GSNR can be derived from the pre-forward-error-correction (FEC) bit-to-error ratio (BER). The GSNR error distribution of DT before and after calibration is shown in Fig. \ref{fig2}(d). The density of the mean square error (MSE) for GSNR prediction is illustrated in Fig. 1(d) using the Gaussian kernel density estimate function. The signs of the errors relative to the ground truth are retained, with positive errors indicating aggressive predictions (larger than the ground truth) and negative errors indicating conservative predictions (lower than the ground truth). Specifically for System 1, an optical spectrum analyzer (OSA) can be used in the laboratory to collect power spectrum. Using these spectrum data, the mean error of GSNR predicted by the DT can be reduced from 1.6 dB to 0.4 dB with refined physical parameters. The error variance is reduced from around 4dB to 0.58dB. In System 2, the topology is much more complex with more unidentified physical parameters. In this case, relying on the total power detected at each amplification site and power profiles detected by the OCM at ROADM sites, the mean GSNR error for up to 12 CUT is reduced from 2.9 dB to 0.8 dB and the error variance is reduced from around 20dB to 2dB. For the deployed C+L-band link of System 3, the mean GSNR error is reduced from 1.8 dB to 0.6 dB and the error variance is reduced from around 0.8dB to 0.5dB. It should be noted that the DT can be continuously updated with collected monitoring data on these three systems. 

\section{LLM-empowered AI Agent for optical network automation}
To develop an LLM-driven AI Agent tailored for autonomous optical networks, it is essential to enable LLMs to access domain-specific knowledge and invoke plugins and tools. In this work, we select GPT-4 as the cognitive Agent for our case study due to its exceptional performance, extensive learning resources, and user-friendly API. However, the LLMs cannot effectively analyze optical network data by simply providing the data and posing questions like "What do these data represent?" or "Can you conclude the system margin from these data?" This is because general-purpose LLMs lack organized and readily utilizable knowledge of optical networks, and can only analyze these data based on its literal meaning. As illustrated in Fig. \ref{fig3}, by integrating domain knowledge using prompt engineering, LLM can conduct effective "\textit{ONet data analysis}". This knowledge encompasses various aspects such as optical components, lightwave systems, fiber nonlinearity, and more, which are categorized and stored in vector form for prompt engineering. The process of prompt engineering consists of several key components including task instructions, chain of thought (CoT), and examples. CoT prompting techniques are employed to guide LLMs through a step-by-step reasoning process, thereby mitigating the limitations posed by insufficient reasoning abilities \cite{wei2022chain}. After such prompt engineering, when analyzing optical network data, the LLM can utilize this domain knowledge to conduct effective analysis and provide helpful insights.

Simultaneously, various tools for addressing different tasks relevant to optical networks have been integrated, including algorithms of the \textit{EDFA configuration optimization }\cite{song2022efficient}, \textit{RSA allocation}, and \textit{provisioning}, as depicted in Fig. \ref{fig3}. These tools are clearly labeled, detailing their usage procedures and adaptation scenarios. Moreover, their input and output data formats are standardized, classifying them as external tools that can be called upon as needed to support the LLM. For example, the most used algorithms of EDFA configuration optimization require inputs such as launch power profiles, number of spans, fiber characteristics, EDFA settings, and other impairments like connector loss. This optimization process involves three steps to adjust the EDFA configuration: first, balancing linear and nonlinear noise; second, countering the effects of Stimulated Raman Scattering (SRS); and third, improving the received optical performance through gradient descent. When invoked, these tools are selected based on similarity analysis with the given problem. The LLM then evaluates whether to utilize an external tool based on the instructions provided by its label. This process is accelerated by techniques of retrieval augmented generation (RAG). Finally, the LLM converts the task problem into computer instructions and invokes the relevant tools to perform calculations through APIs. With these tools, the LLM can solve various specific tasks in optical networks.

\begin{figure*}[t]
\centering\includegraphics[width=1\textwidth]{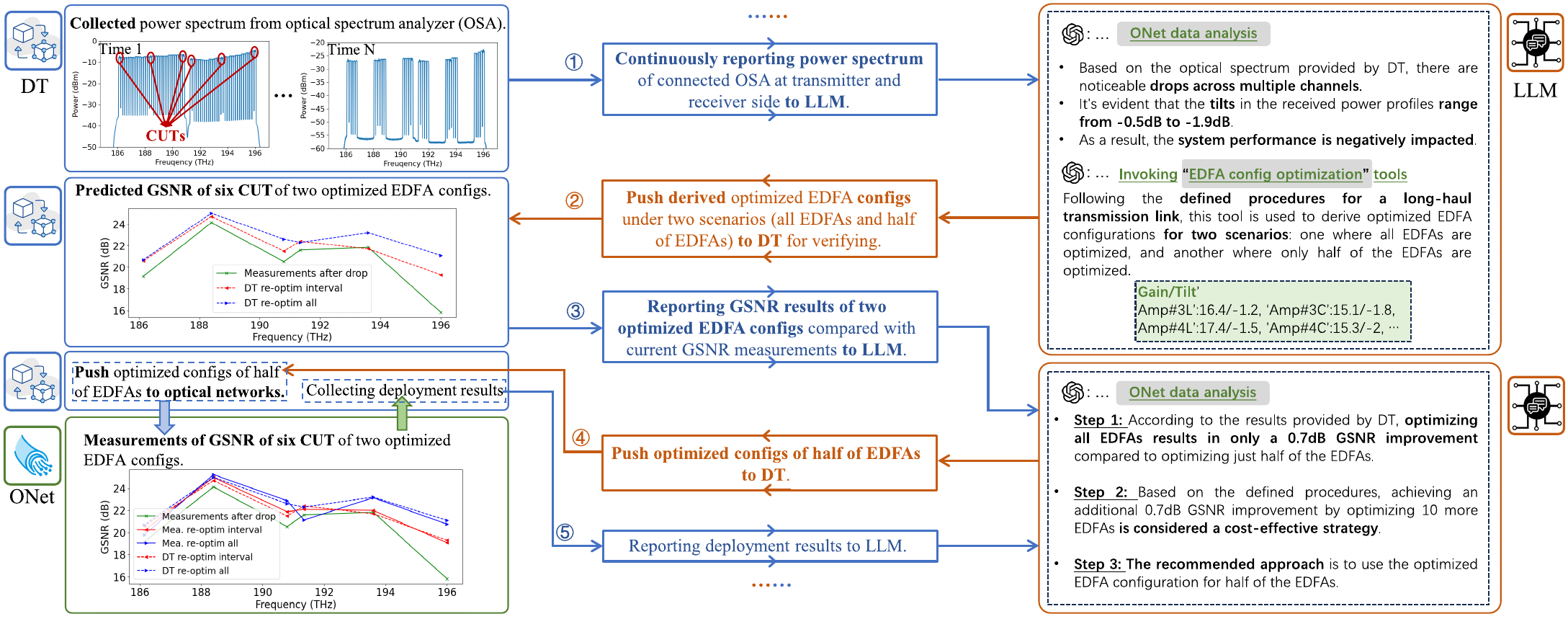}
\caption{Demonstration of autonomous optical performance optimization with dynamic loadings in System 1. Circled numbers represent steps.}
\label{fig4}
\end{figure*}

In our demonstrations, the LLM has been equipped with domain knowledge, plugins, and tools, enabling direct interplay with established DT, as illustrated in Fig. \ref{fig3}. Currently, these integrated tools are not specifically designed for LLMs, and the LLM can only invoke them by analyzing the similarity of human-attached labels. Additionally, the number of integrated tools is limited and covers only a portion of the application scenarios in optical networks. Ideally, the LLM would analyze the functions of corresponding tools by examining their source codes, thereby avoiding the possibility of inaccurate labels. Moreover, the LLM is expected to integrate these tools to solve more complex scenarios. Leveraging domain knowledge, the LLM can analyze specific data from optical networks and extract valuable insights. Additionally, it can utilize the mathematical and physical prediction capabilities of the DT to further enhance its understanding of the current situation.

\section{Demonstrations and Results}
We showcase the interplay between DT and LLM in three different scenarios with various autonomous tasks on these deployed optical transmission systems. In our demonstrations, the DT and LLM are operated offline with collected data of field-deployed or experimental systems.

In System 1 of the experimental long-haul C32+L32 transmission link, 6 CUTs are located at both ends and the middle of the C- and L-bands, with other channels fully loaded using ASE channels. Maintaining satisfactory optical performance under dynamic loading conditions is crucial, particularly for C+L-band long-haul transmission. In this setup, we simulated a signal drop scenario by manually dropping 16 channels from the transmitter side. In this scenario, the DT-enhanced LLM is expected to detect the performance decrease and push optimized EDFA configurations under the new loading condition to this transmission link after verification on the DT. As illustrated in the interplay of step 1 (\textcircled{1}) of Fig. \ref{fig4}, throughout this process, the DT continuously reported data, including power spectrum, to the LLM. By "\textit{ONet data analysis}", the LLM detected drops of multiple channels and observed the increase in the received power profile tilt. Subsequently, to improve transmission performance after signal drops, the LLM invoked the "\textit{EDFA config optimization}" tools. Following predefined procedures in domain knowledge, when the number of spans exceed 10, the LLM uses this tool to derive optimized EDFA configurations for two scenarios: the first one aims to optimize all EDFAs, while in the second one only half of the EDFAs are optimized and the EDFA config between two optimized EDFAs is fixed. The LLM pushes the optimized EDFA configurations for both scenarios to the DT and receives feedback as depicted in step 2 and 3. \textit{ONet data analysis} reveals a 0.7dB GSNR improvement when optimizing all EDFAs compared to optimizing only half. According to predefined procedures (which require at least a 0.1dB improvement for optimizing EDFA in one span), achieving a 0.7dB GSNR improvement by optimizing 10 additional EDFAs is deemed costly. Therefore, the LLM decides to implement the configuration with half of the EDFAs optimized according to predefined knowledge and pushes this configuration to the DT, which then deploys it to the optical network as in step 4. For comparison, the optimized configuration of all EDFAs was also deployed manually. While the DT demonstrated high accuracy with predictions closely matching the measurements, the optimized configuration of all EDFAs resulted in only minimal improvements compared to optimizing half of the EDFAs. It should be noted that, as shown in Fig. \ref{fig1} and step 5 in Fig. \ref{fig4}, the DT can collect measurements from the optical networks post-deployment and report these results to the LLM for further operations or strategy refinement.

\begin{figure*}[t]
\centering\includegraphics[width=1\textwidth]{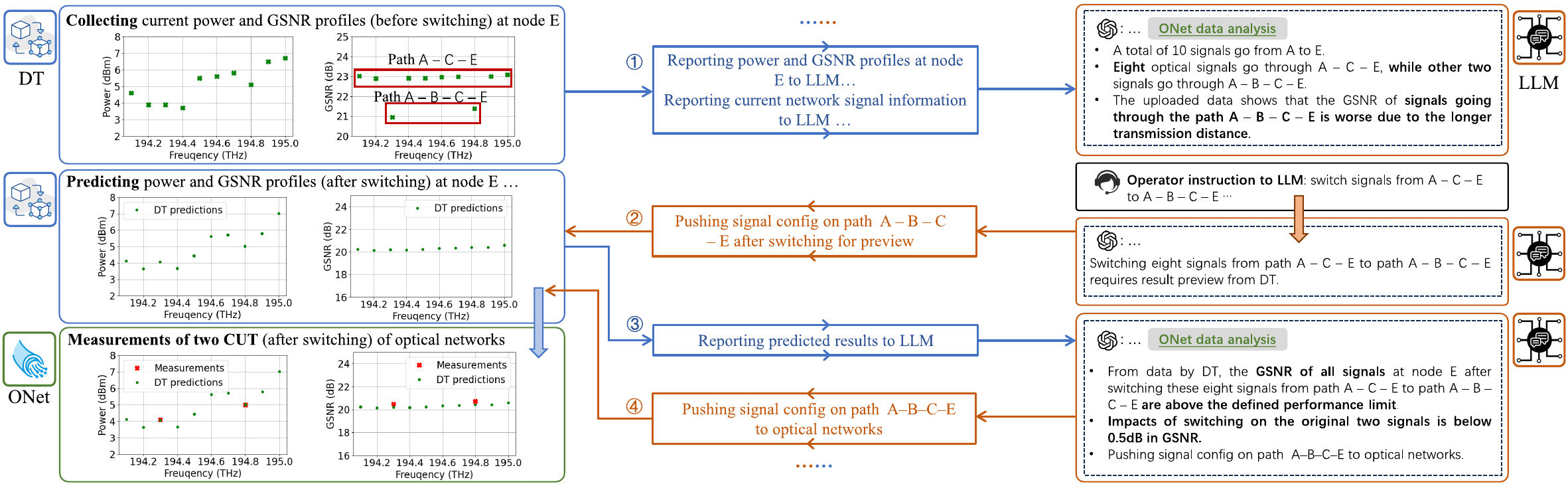}
\caption{Demonstration of autonomous protection switching for device replacement in System 2. Circled numbers represent steps.}
\label{fig5}
\end{figure*}

\begin{figure*}[t]
\centering\includegraphics[width=1\textwidth]{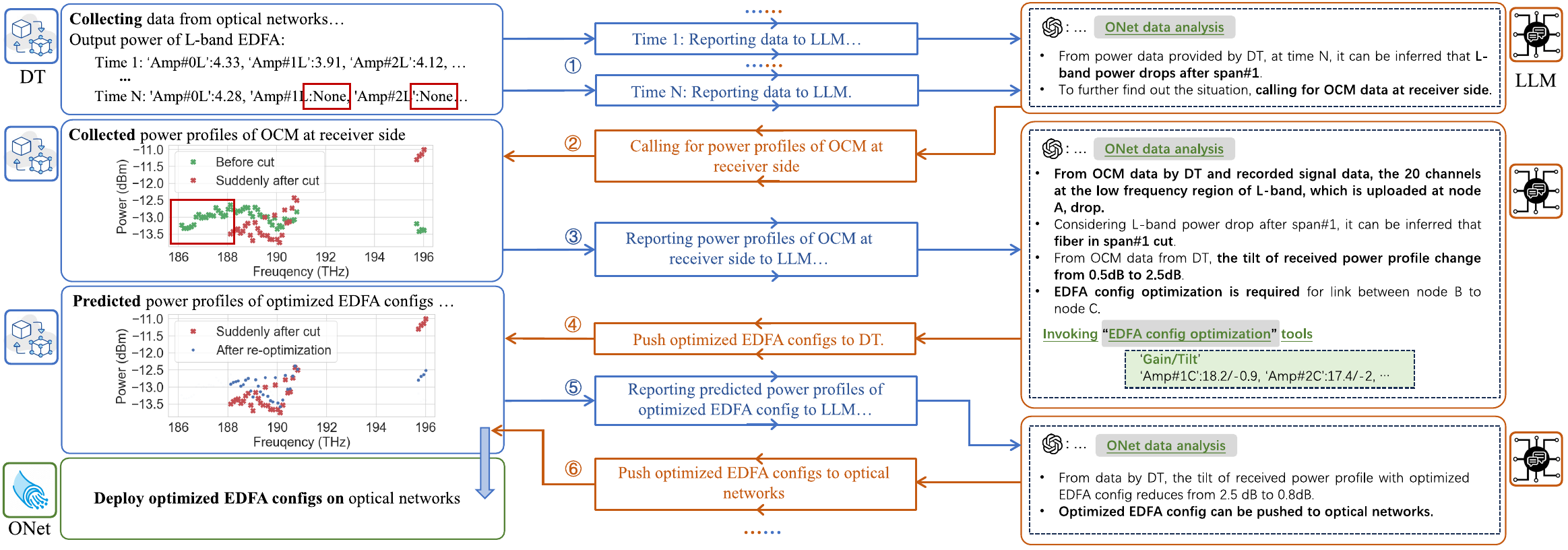}
\caption{Demonstration of autonomous performance recovery after fiber cut in System 3. Circled numbers represent steps.}
\label{fig6}
\end{figure*}

In System 2 of the field-deployed six-node mesh network, we demonstrated a protection switching scenario. Protection switching is a critical and frequently executed step in deployed optical networks for handling device upgrades and component failures, often requiring significant human intervention. The assistance of an LLM is expected to greatly increase efficiency in these processes. Initially, eight optical signals travel through path A-C-E, while two other signals travel through path A-B-C-E. An EDFA is scheduled for replacement on the link from A to C, necessitating the switching of these eight signals to path A-B-C-E. In Fig. \ref{fig5} step 1, this instruction is provided to the LLM by operators. The LLM then pushes the signal configurations for path A-B-C-E to the DT for verification in step 2. Based on the DT's predicted results reported in step 3, the LLM, leveraging its \textit{ONet data analysis} with domain knowledge, derives two useful insights. First, the GSNR of all signals after switching remains above the defined performance limit (18dB in this case). Second, the switching of eight channels has a minimal impact on the original two signals on this path, with GSNR changes below 0.5 dB. Consequently, the LLM pushes the signal configurations for path A-B-C-E to the optical network for deployment as in step 4. As shown in Fig. \ref{fig5}, the measurements of power and GSNR are closely aligned with the DT predictions.

In the System 3 of field-deployed C48+L48 transmission link, the 20 channels at low frequency region of L-band are loaded from node A, while other channels of L-band and four CUTs of C-band are loaded from node B. Transmission link failures, such as fiber cuts, occur frequently and can lead to significant performance degradation. In addition to protection switching, performance optimization is necessary in the downstream links to mitigate the impact of channel drops due to these failures. We simulate a fiber cut scenario, where the fiber of the first span is cut manually, and only signals uploaded from node B remains. During regular operation, the DT continuously reports data to the LLM for analysis. When the fiber is cut, signals uploaded from node A drop. As depicted in step 1 of Fig. \ref{fig6}, at time N, the L-band EDFA after the first span detects no power. To further assess the situation, the LLM requires OCM data at the receiver side in step 2 and step 3. From the drop of the 20 channels of the L-band, the LLM concludes that the first fiber was cut. Moreover, the LLM detects from the OCM data that the tilt of power profiles increases after the fiber cut. In this situation, the LLM autonomously invoked tools of "\textit{EDFA configuration optimization}" and further push optimized configurations to DT for verification as in step 4. Upon obtaining positive results from the DT, these configurations are then pushed to the controllers of the optical networks. 

The three demonstrated scenarios encompass a large portion of the operations in the lifecycle of optical networks, including performance optimization during regular signal load fluctuations, protection switching for device upgrades, and performance recovery after failures. Even without human intervention, the DT-enhanced LLM can solve these problems based on domain knowledge and by invoking corresponding tools, which represents a significant step towards autonomous management.

\section{conclusion}
In this paper, we introduce the framework of a DT-enhanced LLM for autonomous optical networks. LLMs, primarily focused on NLP tasks, are inefficient in simulating and predicting the behaviors of optical networks, which are governed by physical laws. To address this, the accurately updated DT of optical networks provides essential network information to the LLM for decision-making. To specialize the LLM for tasks in optical networks, it is enriched with domain knowledge through prompt engineering and connected to various tools via API integration. Most importantly, strategies derived by the LLM are verified by the DT before deployment to ensure safety. The synergistic interplay between DT and LLM is demonstrated on three deployed optical transmission systems across scenarios involving dynamic loadings, fiber cuts, and protection switching. This approach significantly reduces the need for extensive human intervention in the regular management of optical networks, while enhancing efficiency and maintaining high safety standards.

In future work, beyond offline prototypes, the DT-enhanced LLM will be demonstrated for the online lifecycle management of optical networks. This means that the DT and LLM will operate within the network operative system, adhering to defined protocols. To support the development of an LLM-powered AI agent in optical network management, more tools and plugins with standardized input/output requirements are needed. Additionally, fine-tuning the LLM, rather than relying solely on prompt engineering, will enable it to operate more efficiently within the specific context of optical networks. The DT-enhanced LLM is expected to serve as the cognitive AI Agent, advancing the development of autonomous optical networks.

\section*{Acknowledgments}
The authors would like to acknowledge the China Unicom for providing the field-trial and experimental testbeds for this research.

\vspace{-33pt}
\begin{IEEEbiographynophoto}{Yuchen Song}
(songyc@bupt.edu.cn) is a Ph.D. candidate at the Beijing University of Posts and Telecommunications (BUPT).
\end{IEEEbiographynophoto}
\vspace{-33pt}
\begin{IEEEbiographynophoto}{Yao Zhang}
(zhang-yao@bupt.edu.cn) is a Ph.D. candidate at BUPT.
\end{IEEEbiographynophoto}

\vspace{-33pt}
\begin{IEEEbiographynophoto}{Anni Zhou}
(zhouanni@bupt.edu.cn) is currently pursuing a Master degree at BUPT.
\end{IEEEbiographynophoto}

\vspace{-33pt}
\begin{IEEEbiographynophoto}{Yan Shi}
(shiyan49@chinaunicom.cn) currently works at China Unicom Research Institute. She received her Ph.D. degree from BUPT.
\end{IEEEbiographynophoto}

\vspace{-33pt}
\begin{IEEEbiographynophoto}{Shikui Shen}
(shensk@chinaunicom.cn) currently works at China Unicom Research Institute. He mainly focuses on the technique research and standardization of optical networks.
\end{IEEEbiographynophoto}

\vspace{-33pt}
\begin{IEEEbiographynophoto}{Xiongyan Tang}
(tangxy@chinaunicom.cn) currently works at China Unicom Research Institute. He is the Chief Scientist with China Unicom Research Institute and the Vice Dean of China Unicom Research Institute.
\end{IEEEbiographynophoto}

\vspace{-33pt}
\begin{IEEEbiographynophoto}{Jin Li}
(jinlee@bupt.edu.cn) received his Ph.D. degree from BUPT. He is currently a post-doc researcher at BUPT.
\end{IEEEbiographynophoto}

\vspace{-33pt}
\begin{IEEEbiographynophoto}{Min Zhang}
(mzhang@bupt.edu.cn) is currently a professor with BUPT. His main research interests include advanced optical communication systems and networks.
\end{IEEEbiographynophoto}

\vspace{-33pt}
\begin{IEEEbiographynophoto}{Danshi Wang}
(Senior Member, IEEE) (danshi\_wang@bupt.edu. cn) is currently an associate professor with the Institute of Information Photonics and Optical Communications, Beijing University of Posts and Telecommunications (BUPT). He has published more than 170 articles. His research interests include AI for science, digital twin optical network.
\end{IEEEbiographynophoto}

\vfill

\end{document}